\documentclass[prl,aps,twocolumn,showpacs,preprintnumbers,amsmath,amssymb]{revtex4}
\usepackage{epsfig}
\usepackage{color}
\usepackage{ifthen}
\usepackage{graphicx}
\usepackage{amsmath}
\usepackage{amssymb}

\usepackage{graphicx}
\usepackage{dcolumn}
\usepackage{bm}
\def\Dsl{\hbox{/\kern-.6700em\it D}} 
\def\dsl{\hbox{/\kern-.5300em$\partial$}}

\def\eqa{\begin{eqnarray}}
\def\eeqa{\end{eqnarray}}
\def\eq{\begin{equation}}
\def\eeq{\end{equation}}
\def\be{\begin{equation}}
\def\ee{\end{equation}}
\def\bea{\begin{eqnarray}}
\def\eea{\end{eqnarray}}

\newcommand{\dslash}{\not{\hbox{\kern-2pt $\partial$}}}
\newcommand{\pslash}{\not{\hbox{\kern-2.3pt $p$}}}
 \newtoks\nslashfraction
 \nslashfraction={.13}
 \newcommand{\nslash}[1]{\setbox0\hbox{$ #1 $}
   \setbox0\hbox to \the\nslashfraction\wd0{\hss \box0}/\box0 }

\begin{document}

\preprint{}

\title{Angular Correlation Functions for Models with Logarithmic Oscillations}
\author{Mark G. Jackson$^{1,2,3}$, Ben Wandelt$^{1,2}$, Fran\c{c}ois Bouchet$^1$}
\affiliation{$^1$Institut d'Astrophysique de Paris, UMR CNRS 7095,}
\affiliation{Universit\'{e} Pierre et Marie Curie, 98bis boulevard Arago, 75014 Paris, France}
\affiliation{$^2$Department of Physics, University of Illinois at Urbana-Champaign, Urbana, IL 61801}
\affiliation{$^3$Paris Centre for Cosmological Physics and Laboratoire AstroParticule et Cosmologie, \\Universit\'{e} Paris 7-Denis Diderot, Paris, France}

\date{\today}


\begin{abstract}
\noindent
There exist several theoretical motivations for primordial correlation functions (such as the power spectrum) to contain oscillations as a logarithmic function of comoving momentum $k$.  While these features are commonly searched for in $k$-space, an alternative is to use angular space; that is, search for correlations between the directional vectors of observation.  We develop tools to efficiently compute the angular correlations based on a stationary phase approximation and examine several example oscillations in the primordial power spectrum, bispectrum, and trispectrum.  We find that logarithmically-periodic oscillations are essentially featureless and therefore difficult to detect using the standard correlator, though others might be feasible.
\end{abstract}

\maketitle
\section{Introduction}
The leading paradigm in modern cosmology is inflationary theory, in which quantum field vacuum energy causes the Universe to undergo a very rapid expansion in a very short amount of time \cite{Guth:1980zm,Linde:1981mu,Albrecht:1982wi,Linde:1983gd}.  One of the consequences of inflation is that the quantum field fluctuations source structure formations, which then become fluctuations in the CMB temperature and polarization as well as large scale structure.  Thus, precision cosmological measurements will yield tight constraints on the quantum field interactions, and hence the microscopic details of the inflationary theory.  Such measurements are now being made with increasing precision by \emph{COBE} \cite{Smoot:1992td} and \emph{WMAP} \cite{Hinshaw:2012fq,Bennett:2012fp}, and there will soon be data from the \emph{Planck} \cite{planck} and \emph{Euclid} \cite{euclid} satellites, and suborbital or ground-based polarization-dedicated experiments \cite{cmbpol}.

The simplest inflation model, and the one considered here, contains a single field $\phi$ which can be decomposed into the background and fluctuations as
 \[ \phi( {\bf x}, \tau) = \phi_0 (\tau) + \varphi ( {\bf x},\tau) . \]
We use the conformal time $-\infty~<~\tau~<~0^-$ such that $\tau \rightarrow 0^-$ represents future infinity, and then Fourier transform the fluctuations into comoving momentum ${\bf k}$-space modes $\varphi_{\bf k}(\tau)$.  
The interactions of the underlying inflationary quantum field theory are then encoded in the correlation functions of $\varphi_{\bf k}$.  It is these microscopic interactions which determine the quantum field theory model, and hence the precise mechanism responsible for cosmological inflation.  While a variety of inflationary models have been proposed, differentiation between them requires precision measurements of possible features in the correlation functions.  Such a measurement is then tantamount to explaining how the Universe came into being.

While some of the correlations may contain features slowly varying as a function of $k$, others could be sharper.  One such feature would be oscillations.  This modulation could occur through high-energy physics encoded in the choice of vacuum \cite{Brandenberger:1999sw,Greene:2005aj, Kempf:2000ac, Niemeyer:2001qe, Kempf:2001fa, Martin:2000xs, Niemeyer:2000eh, Brandenberger:2002hs, Martin:2003kp, Easther:2001fi, Easther:2001fz, Easther:2002xe, Kaloper:2002uj, Kaloper:2002cs, Danielsson:2002kx, Danielsson:2002qh, Shankaranarayanan:2002ax, Hassan:2002qk, Bozza:2003pr, Alberghi:2003am,Schalm:2004xg,arXiv:1007.0185, arXiv:1104.0887, Jackson:2012fu, Jackson:2012qp}, sharp turns in multifield inflation models \cite{Achucarro:2010jv, Achucarro:2010da, Achucarro:2012sm, Achucarro:2012yr, Shiu:2011qw,Gao:2012uq}, or quasi-periodic backgrounds \cite{Falciano:2008gt,Flauger:2009ab,Chen:2008wn}.  While most attention has focused on oscillations in the power spectrum, there has also been increasing interest in the bispectrum \cite{Meerburg:2009ys, Meerburg:2009fi, Flauger:2010ja, Meerburg:2010ca} and trispectrum \cite{Jackson:2012fu}.

This developing understanding of the relationship between fundamental physics and primordial oscillations motivates an efficient method to detect such features.  Such searches typically have analyzed the correlation function as a function of $k$-space \cite{Martin:2003sg, Martin:2004iv, Martin:2004yi, Easther:2004vq, Easther:2005yr,Meerburg:2011gd,Peiris:2013opa}.  This comparison is usually done by transforming the primordial perturbation into the CMB temperature fluctuation at degree~$\ell$.  In the approximation of instantaneous CMB decoupling at time~$\eta$ the transformation is given by 
\[  \frac{\Delta T_\ell}{T} \equiv \int \frac{ d^3 {\bf k}}{(2 \pi)^3} \ \varphi_{\bf k}(0^-) T_{\rm rad}(k) j_\ell( \eta k) \]
where $T_{\rm rad}(k)$ is the radiative transfer function.  Here we have simplified the gauge-fixing and freeze-out issues, which are irrelevant to the discussion at hand.  Each $\ell$-mode receives contributions from angular separation $\delta \theta \approx \pi/\ell$.  Correlations of $\Delta T_\ell/T$ can then be evaluated using correlations of $\varphi_{\bf k}$, such as the power spectrum
\[ C_\ell \equiv \langle \left( \frac{\Delta T_\ell}{T} \right)^2 \rangle . \]
While this is indeed the natural way to constrain slight scale-dependence, it is not so natural for detecting oscillations.  

\newpage
A more promising approach for detecting oscillations would be to Fourier transform to angular, or direction, space \cite{Bashinsky:2000uh} via
\begin{equation}
\label{dirFourier}
 \frac{\Delta T({\hat {\bf n}})}{T} = \int \frac{ d^3 {\bf k}}{(2 \pi)^3} \ \varphi_{\bf k}(0^-) T_{\rm rad}(k) e^{-i \eta {\bf k} \cdot {\hat {\bf n}}} .
\end{equation} 
This has the advantage that the observation is completely localized to a single direction, and so one no longer needs to sample many region of the sky to gain information for a single $\ell$-mode.  This may also produce a stronger signal and allow a more efficient computational processing.  Such an approach has not yet been widely explored, and is well worth investigating.

In this article we calculate the angular correlation functions for models containing oscillations.  We focus upon oscillations which are periodic in the logarithm of $k$ since these are physically motivated by new physics arising at energy scale $M$.  We use a stationary phase approximation which takes advantage of the oscillatory features.  From this we derive the dominant configurations of the direction-vectors and explicitly evaluate the correlation function at these locations.  The result is correct to leading order in $H/M$, where $H$ is the Hubble parameter determining the energy scale of inflation. To maximize the possible benefit of our new approach we assume instantaneous decoupling; were we to relax this assumption, oscillations more rapid than the timescale of decoupling would become smeared out.  In \S1 we evaluate an example in the primordial power spectrum.  In \S2 we apply this technique to an example in the bispectrum, and in \S3 we evaluate the trispectrum.  In \S4 we discuss the results and offer observational prospects.

\section{Power Spectrum}
\begin{figure}
\begin{center}
\includegraphics[width=3in]{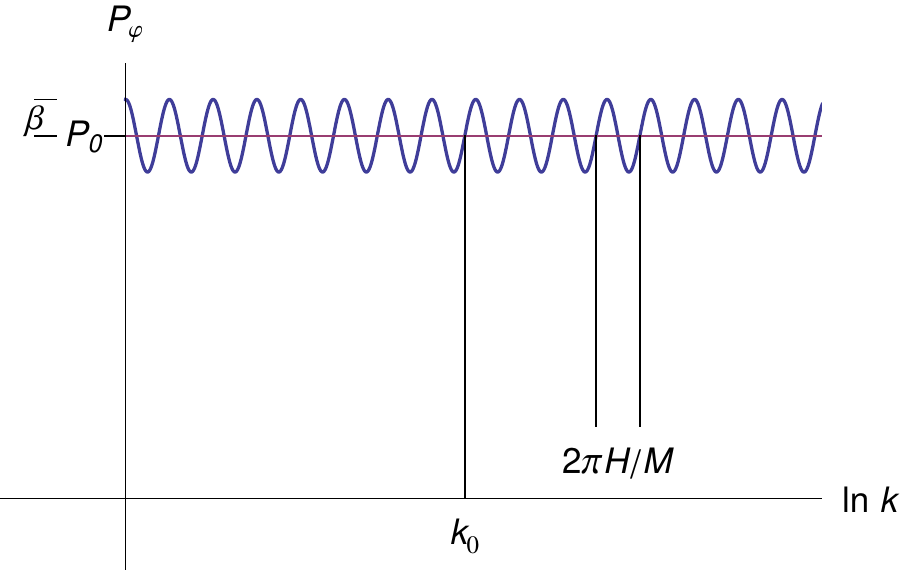}
\caption{Primordial power spectrum containing oscillations periodic in $\ln k$.  For reference, the scale-invariant ``flat" power spectrum is given by the thin red line.}
\label{pskspace}
\end{center}
\end{figure}

Let us begin with the power spectrum, or two-point correlation function:
\[  \langle  \varphi_{\bf k}(0^-) \varphi_{\bf k'}(0^-) \rangle \equiv \frac{2 \pi^2}{k^3} P_\varphi(k)  (2 \pi )^3 \delta^3 ( {\bf k} + {\bf k}') . \]
Suppose the power spectrum is mostly scale-invariant but contains oscillations set by the scale $H/M$, which is strongly theoretically motivated \cite{Brandenberger:1999sw,Greene:2005aj, Kempf:2000ac, Niemeyer:2001qe, Kempf:2001fa, Martin:2000xs, Niemeyer:2000eh, Brandenberger:2002hs, Martin:2003kp, Easther:2001fi, Easther:2001fz, Easther:2002xe, Kaloper:2002uj, Kaloper:2002cs, Danielsson:2002kx, Danielsson:2002qh, Shankaranarayanan:2002ax, Hassan:2002qk, Bozza:2003pr, Alberghi:2003am,Schalm:2004xg,arXiv:1007.0185, arXiv:1104.0887, Jackson:2012fu, Jackson:2012qp}:
\begin{equation}
\label{pphi}
P_\varphi(k) = P_0 \left[ 1 + \beta \cos \left(  \frac{M}{H} \ln \frac{k}{k_0} \right) \right]
 \end{equation}
 where $k_0$ is some reference wavenumber and $\beta$ is a small parameter controlling the oscillation magnitude, as shown in Figure~\ref{pskspace}.  It was shown in \cite{markgary} that there is a theoretical bound on the oscillation frequency.

\begin{figure}
\begin{center}
\includegraphics[width=3in]{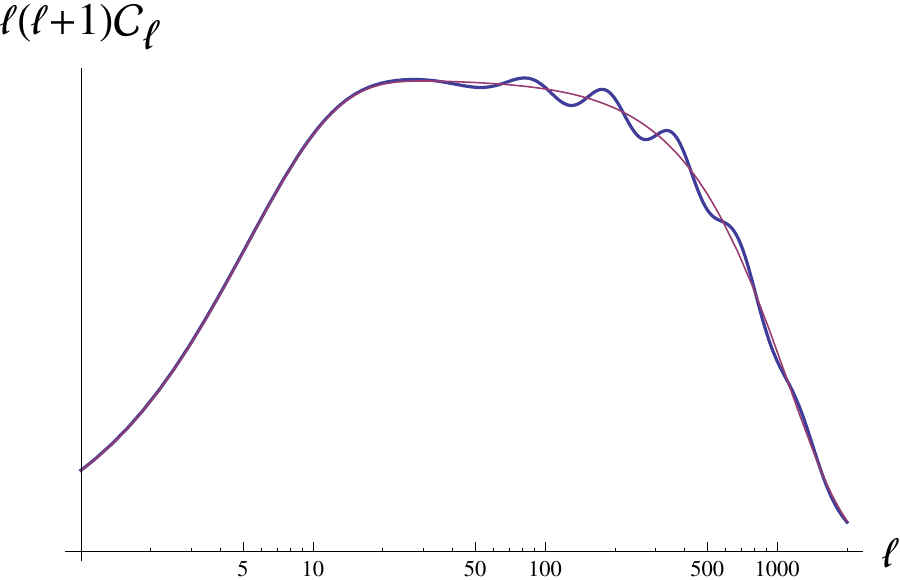}
\caption{Simplified CMB temperature power spectrum in the $\ell$-basis.  The flat power spectrum is given by the thin red line.}
\label{psellspace}
\end{center}
\end{figure}

A quantity of great interest is the CMB temperature fluctuation power spectrum,
\begin{eqnarray}
\label{cl}
C_\ell &\equiv& \langle  \frac{\Delta T_\ell}{T}  \frac{\Delta T_\ell}{T}  \rangle \\
\nonumber
 &=& \int \frac{ d^3 {\bf k}}{(2 \pi)^3}  \frac{ d^3 {\bf k}'}{(2 \pi)^3} \langle  \varphi_{\bf k} \varphi_{\bf k'} \rangle T_{\rm rad}(k)  j_\ell( \eta k) T_{\rm rad}(k') j_\ell (\eta k') .
 \end{eqnarray}
The transfer function $T_{\rm rad}(k)$ is in general a complicated function of scale and cosmological parameters, but at both large and small angular scales it can be approximated as $T_{\rm rad}(k) \approx e^{-k^2/k_D^2}$ for some effective diffusion scale $k_D$.  For simplicity of an analytic answer we will use this function, and the acoustic peaks appearing at intermediate scales would be straightforward to include numerically.   Evaluating this with the primordial oscillation (\ref{pphi}) gives the power spectrum shown in Figure~\ref{psellspace}, demonstrating that oscillations now appear as (damped) oscillations in the $C_\ell$.

An alternative, though much less common, representation is to evaluate the temperature fluctuation power spectrum in position space,
\begin{eqnarray*}
 && \hspace{-0.2in} \langle  \frac{\Delta T({\hat {\bf n}}_1)}{T}  \frac{\Delta T({\hat {\bf n}}_2)}{T}  \rangle \\
 &=& \int \frac{ d^3 {\bf k}}{(2 \pi)^3}  \frac{ d^3 {\bf k}'}{(2 \pi)^3} \langle  \varphi_{\bf k} \varphi_{\bf k'} \rangle T_{\rm rad}(k) e^{-i \eta {\bf k} \cdot {\hat {\bf n}}_1} T_{\rm rad}(k') e^{ -i \eta {\bf k}' \cdot {\hat {\bf n}}_2 } . 
 \end{eqnarray*}
We now define the rescaled conformal time $\gamma~\equiv~H \eta/M$, and use the unit vector difference
\[ {\bf N} \equiv  {\hat {\bf n}}_1 - {\hat {\bf n}}_2 \]
so $0 \leq N \leq 2$.  Due to isotropy, the power spectrum can only depend on $N$.  Writing the cosine as exponentials, the correction term can now be cast into the form
\[  \left( \frac{\Delta T}{T}  \right)^2_{\rm osc} = \int \frac{ d^3 {\bf k}}{(2 \pi)^3} G( {\bf k}) \left[ e^{-i \frac{M}{H} F_+({\bf k})} + e^{-i \frac{M}{H} F_-({\bf k})} \right] \]
where
\begin{eqnarray*}
G( {\bf k}) &=& \frac{ \pi^2 \beta P_0}{k^3} e^{-2 k^2/k_D^2}, \\ 
 F_\pm({\bf k})  &=& \pm \ln \frac{k}{k_0} + \gamma {\bf k} \cdot {\bf N} . 
 \end{eqnarray*}
The majority of the contribution will come from near the extremum ${\bf k}_{*\pm}$ at
\[ 0 = \frac{\partial F_\pm}{\partial k^a} = \pm \frac{k^a_{*\pm}}{k^2_{*\pm}} + \gamma N^a. \]
This is easily solved to yield
\[ {\bf k}_{*\pm} = \mp \frac{ {\bf N}}{\gamma N^2} . \]
At this moment of stationary phase,
\begin{eqnarray*}
G( {\bf k}_{*\pm}) &=&  \pi^2 \beta P_0 (\gamma N)^3 e^{- 2(N \gamma k_D)^{-2}}, \\
F_\pm( {\bf k}_{*\pm}) &=& \mp \left[ \ln (\gamma N k_0) + 1 \right]
\end{eqnarray*}
and the fluctuations around this point are given by
\begin{eqnarray*}
\mathcal M^{ab}_\pm &\equiv& \left.  \frac{\partial^2 F_\pm}{\partial k^a \partial k^b} \right|_{{\bf k}_*} \\
&=& \pm \gamma^2  \left( N^2 \delta^{ab} - 2 N^a N^b \right).
\end{eqnarray*}
\begin{figure}
\begin{center}
\includegraphics[width=3in]{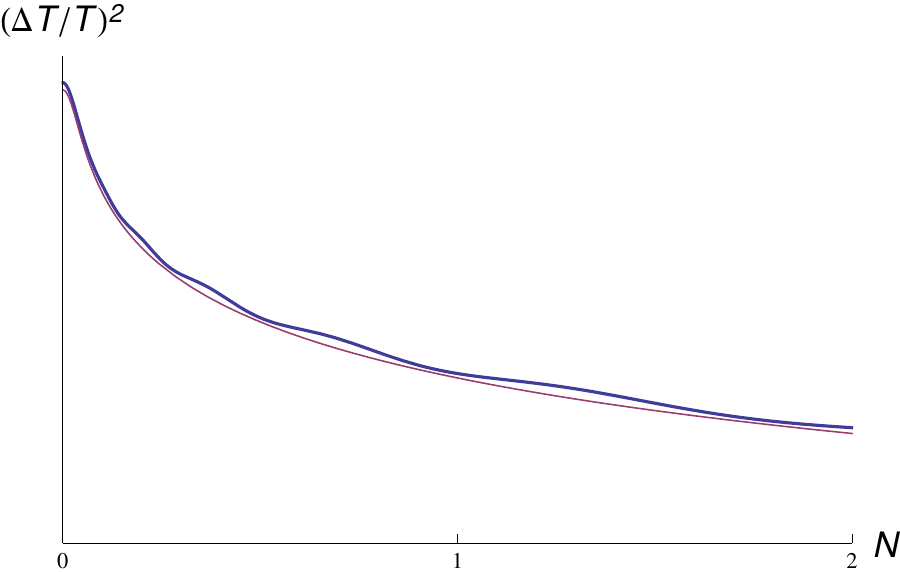}
\caption{Simplified CMB temperature power spectrum in the direction-basis, as a function of directional separation $N$.  The flat power spectrum is given by the thin red line.}
\label{psdirspace}
\end{center}
\end{figure}
After performing the Gaussian integral the answer is then conveniently written as
\begin{eqnarray*}
\left( \frac{\Delta T}{T}  \right)^2_{\rm osc} &\approx& G(k_{*+}) e^{-i \frac{M}{H} F_+( {\bf k}_{*+})} \\
&& \hspace{-0.0in} \times \int \frac{d^3 {\bf k}}{(2 \pi)^3} e^{-i \frac{M}{2H} (k^a-k^a_{*+}) (k^b-k^b_{*+}) \mathcal M^{ab}_+ } + {\rm c.c.} \\
&=& G(k_{*+}) e^{-i \frac{M}{H} F_+({\bf k}_{*+})} \sqrt{ \frac{(iH)^3}{ (2 \pi) ^{3} M^3 \left| \mathcal M_+ \right|} } + {\rm c.c.}
\end{eqnarray*}
To explicitly evaluate this we use rotational invariance to choose coordinates such that
\[ {\bf N} = N {\hat {\bf z}} \]
which makes 
\[ \left| \mathcal M_{\pm} \right| =  (\gamma N)^6. \]
The final answer for the oscillation term is then 
\begin{eqnarray}
\nonumber
\left( \frac{\Delta T}{T}  \right)^2_{\rm osc} &=& 2 \pi^2 \beta P_0  \left( \frac{H}{2 \pi M} \right)^{3/2} e^{- 2(N \gamma k_D)^{-2}} \\
\label{finalspectrum}
&& \times \cos \left( \frac{M}{H} \left[ \ln (\gamma N k_0) + 1 \right] \right) . 
\end{eqnarray}
This is shown in Figure~\ref{psdirspace}.  Contrary to the expectation that there would be a single large peak, there are again multiple oscillations.  This is due to the primordial power spectrum being logarithmically periodic, which dilutes the signal over all scales.  We will find this behavior continue to higher-point correlators.

\section{Bispectrum}
We now consider a more elaborate calculation, the position-space primordial bispectrum:
\begin{eqnarray*}
&& \hspace{-0.2in} \langle  \frac{\Delta T({\hat {\bf n}}_1)}{T}  \frac{\Delta T({\hat {\bf n}}_2)}{T}  \frac{\Delta T({\hat {\bf n}}_3)}{T}   \rangle \\
\nonumber
&=&  \int \prod_{i=1}^3 \frac{ d^3 {\bf k}_i}{(2 \pi)^3} T_{\rm rad}(k_i) \exp \left( - i \eta  {\bf k}_i  \cdot {\hat {\bf n}}_i \right) \langle \varphi_{{\bf k}_1} \varphi_{{\bf k}_2} \varphi_{{\bf k}_3} \rangle
\end{eqnarray*}
where the correlation is of the form
\[ \langle \varphi_{{\bf k}_1} \varphi_{{\bf k}_2} \varphi_{{\bf k}_3} \rangle  \equiv B_\varphi( {\bf k}_1, {\bf k}_2, {\bf k}_3)  (2 \pi)^3 \delta^3 \left( \sum_{i=1}^3 {\bf k}_i  \right) . \]
Since ${\bf k}_1+{\bf k}_2+{\bf k}_3=0$ the $k$-space correlations are categorized by the triangle formed by the ${\bf k}_i$.  Oscillations (or other features) could then depend arbitrarily on both the size and shape of the triangle.  Let us consider the motivation for these two types of bispectrum oscillations.

The primordial fluctuations have been found to be very nearly scale-invariant.  Though oscillations in the power spectrum (by definition) break this invariance, it is possible to maintain perfect scale-independence and yet have oscillations in the \emph{bispectrum}.  This could be accomplished via a phase which depends only upon the relative angles formed by the ${\bf k}_i$ and not by their overall magnitude, being invariant under $k_i \rightarrow \lambda k_i$.  Despite this strong theoretical motivation, it appears difficult to formulate any non-trivial scale-independent bispectrum oscillations which allow a stationary phase approximation.  Thus here we will limit ourselves to scale-dependent oscillations.  These would contribute to other scale-dependent features in the bispectrum, and which are possibly measurable in the near future \cite{Sefusatti:2009xu}.

\subsection{Scale-Dependent Oscillations}
In slow-roll inflation the bispectrum induced by gravitational interactions is negligibly small \cite{Maldacena:2002vr} so we can focus exclusively on the term generated by field interactions of a particular model.  Consider the example
\begin{equation}
\label{bi1}
B_\varphi  = \frac{ B_0}{(k_1 k_2 k_3)^{2}} \left( \frac{k^2_1 + k^2_2 + k^2_3}{3k_0^2} \right)^{-i \frac{M}{2H} }  + {\rm c.c.}
 \end{equation}
This is similar to the ``Resonant Non-Gaussianity" model studied in \cite{Flauger:2010ja}.  As in the previous power spectrum example, the $(k_1 k_2 k_3)^{-2}$ compensates for the phase-space factor; the bispectrum (\ref{bi1}) then depends only on the sum $\sum_i k_i^2$, and so is well-behaved for any non-vanishing triangle.

In order to perform the integrals over ${\bf k}_i$ we first employ the Fourier representation of the momentum-conserving delta-function, where we include a factor of $M/H$ for future convenience:
\begin{equation}
\label{deltafunct}
 (2 \pi)^3 \delta^3 \left( \sum_{i=1}^3 {\bf k}_i  \right) =  \left( \frac{M}{H} \right)^3 \int d^3 {\bf w} \exp - i \frac{M}{H} \left( \sum_{i=1}^3 {\bf k}_i \right) \cdot {\bf w} . 
\end{equation}

Performing the same rescaling as before, the functions are now 
\begin{eqnarray}
\label{biaction}
G( {\bf k}) &=& \frac{B_0}{ (k_1 k_2 k_3)^{2}} e^{- \sum_{i=1}^3 k^2_i /k_D^2}, \\ 
\nonumber
F_\pm( {\bf k}_i ,{\bf w}) &=&\pm \frac{1}{2}  \ln \frac{k^2_1 + k^2_2 + k^2_3}{3k_0^2} \\
\nonumber
&+& \gamma \sum_{i=1}^3 {\bf k}_i \cdot {\hat {\bf n}}_i + {\bf w} \cdot \left( \sum_{i=1}^3 {\bf k}_i \right) .
 \end{eqnarray}

The extremum will satisfy
\[ 0 = \frac{\partial F_\pm}{\partial k_i^a} = \pm \frac{k_{*\pm,i}^a}{k^2_{*\pm,1} + k^2_{*\pm,2} + k^2_{*\pm,3}} + \gamma {\hat {n}}_i^a + w^a. \]
For now we will treat ${\bf w}$ as a constant and solve for ${\bf k}_{*\pm,i}$ in terms of it.  This yields
\begin{equation}
\label{bisol}
 {\bf k}_{*\pm,i} = \mp \frac{\gamma {\hat {\bf n}}_i + {\bf w}}{  \sum_i \left| \gamma {\hat {\bf n}}_i + {\bf w} \right|^2 } . 
 \end{equation}
The ${\bf k}_i$-fluctuation matrix near this point is
\begin{eqnarray*}
\mathcal M^{{\bf k},ab}_{\pm,ij} &\equiv& \left. \frac{\partial^2 F_\pm}{\partial k^a_i \partial k^b_j} \right|_{{\bf k}_*} \\
&& \hspace{-0.5in} = \pm \left(  \sum_k \left| \gamma {\hat {\bf n}}_k + {\bf w} \right|^2 \right) \delta_{ij} \delta^{ab} \mp 2\left( \gamma {\hat { n}}^a_i + w^a \right) \left( \gamma {\hat { n}}^b_j + w^b \right) .
\end{eqnarray*}

Substituting the solution (\ref{bisol}) back into the exponent $F_\pm$ in (\ref{biaction}) yields
\[ F_\pm( {\bf w}) = \mp  \left[ \frac{1}{2} \ln \left( 3k_0^2 \sum_i \left| \gamma {\hat {\bf n}}_i + {\bf w} \right|^2 \right) + 1 \right]. \]
We then treat ${\bf w}$ as another parameter to vary, whose extremum yields the momentum conservation equation
\begin{equation}
\label{biext2}
 0 = \frac{\partial F_\pm}{\partial w^a} = \mp \frac{ \sum_{i=1}^3 \gamma {\hat { n}}^a_i + { w}^a_* }{\sum_j \left| \gamma {\hat {\bf n}}_j + {\bf w}_* \right|^2 }. 
 \end{equation}
Note that we could have obtained this by taking $\partial F_\pm / \partial w^a$ in (\ref{biaction}) originally, but obtaining it in this order makes evaluation of the fluctuation-matrix determinant much easier.  The solution to (\ref{biext2}) is 
\[ {\bf w}_* = - \frac{\gamma}{3} \sum_i {\hat {\bf n}}_i. \]
The fluctuations around this point are
\begin{eqnarray*}
\mathcal M^{{\bf w},{ab}}_\pm &\equiv& \left. \frac{ \partial^2 F_\pm}{\partial w^a \partial w^b} \right|_{{\bf k}_{*,i}, {\bf w}_*} \\
&=& \pm \gamma^{-2} \left[ 3 - \frac{1}{3} \left( \sum_i {\hat {\bf n}}_i \right)^2 \right]^{-1} \delta^{ab} . \\
\end{eqnarray*}
The fact that $\mathcal M^{\bf w} \sim \left( \mathcal M^{\bf k} \right)^{-1}$ fits our intuition that ${\bf w}$ is a ``negative degree of freedom," constraining the ${\bf k}_i$ to be conserved.  

The temperature bispectrum is then
\begin{eqnarray}
\nonumber
&& \langle \frac{\Delta T({\hat {\bf n}}_1)}{T}  \frac{\Delta T({\hat {\bf n}}_2)}{T}  \frac{\Delta T({\hat {\bf n}}_3)}{T}   \rangle \\
\nonumber
&=& G( {\bf k}_*)e^{- i \frac{M}{H} F_+({\bf w}_* )} \left( \frac{M}{H} \right)^3 \int d^3 {\bf w} \int \prod_{i=1}^3 \frac{ d^3 {\bf k}_i}{(2 \pi)^3} \\
\nonumber
&& \hspace{-0.3in} \times e^{-i \frac{M}{2H} (k^a_i-k^a_{*+,i}) (k^b_j-k^b_{*+,j}) \mathcal M^{{\bf k},ab}_{+,ij} - i \frac{M}{2H} (w^a-w^a_{*}) (w^b-w^b_{*})  \mathcal M^{{\bf w},ab}_+} + {\rm c.c.} \\
\nonumber
&=& \frac{B_0}{ (2 \pi)^3 } \left( \sum_i \left| \gamma {\hat {\bf n}}_i + {\bf w}_* \right|^2 \right)^{6} \prod_{j=1}^3 \frac{e^{- (| \gamma {\hat {\bf n}}_j + {\bf w}_* |  k_D)^{-2}}}{| \gamma {\hat {\bf n}}_j + {\bf w}_* |^{2}}  \\
\label{finalbi}
&&\hspace{0.4in} \times \left( \frac{H}{M} \right)^3 \frac{e^{- i \frac{M}{H} F_+({\bf w}_* )} }{\sqrt{\left| \mathcal M^{{\bf k}}_+ \right| \left| \mathcal M^{{\bf w}}_+ \right|} } + {\rm c.c.} 
 \end{eqnarray}
where
\[  F_\pm({\bf w}_* ) = \mp \left(  \ln \left[ \gamma k_0 \sqrt{ 9 - \left( \sum_i {\hat {\bf n}}_i \right)^2 } \right] + 1 \right). \]

\subsection{A Family of Solutions}
\begin{figure}
\begin{center}
\includegraphics[width=1.3in]{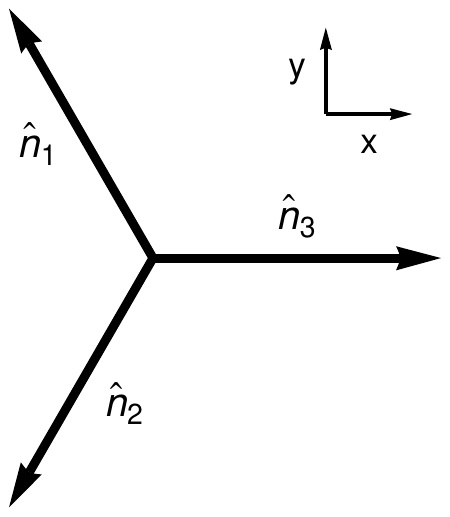}
\hspace{0.1in}
\includegraphics[width=1.3in]{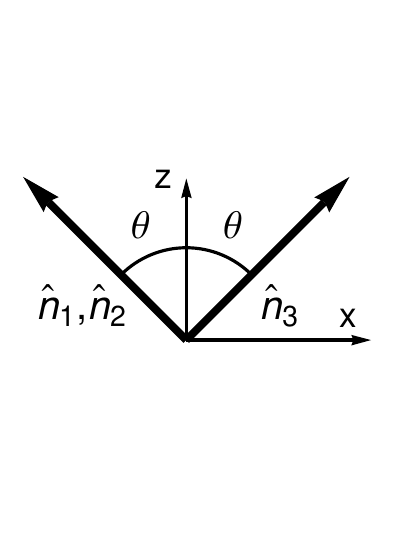}
\ \\
\includegraphics[width=1.5in]{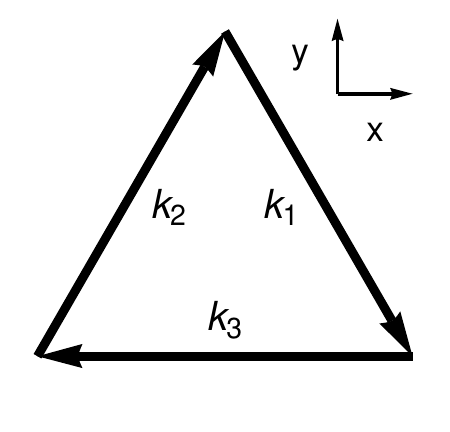}
\caption{The directional vector configuration (top) resulting in a $k$-space equilateral triangle (bottom).}
\label{allbi}
\end{center}
\end{figure}

While analyzing the general bispectrum solution (\ref{finalbi}) is difficult, it is simple to consider a one-parameter family in which the symmetric directional vectors all make an angle $\theta$ with some ${\hat {\bf z}}$-axis as shown in Figure~\ref{allbi}:
\[ {\hat {\bf n}}_i = \sin \theta \cos \frac{2 \pi i}{3} {\hat {\bf x}} +  \sin \theta \sin \frac{2 \pi i}{3} {\hat {\bf y}} +  \cos \theta {\hat {\bf z}}. \]
This makes the solution
\begin{eqnarray*}
 {\bf w}_* &=& - \gamma \cos \theta, \\
 {\bf k}_{*,i} &=& - \frac{ \left( \cos \frac{2 \pi i}{3} {\hat {\bf x}} + \sin \frac{2 \pi i}{3} {\hat {\bf y}} \right)}{  3 \gamma \sin \theta } , \\
 | \gamma {\hat {\bf n}}_i + {\bf w}_* | &=& \gamma \sin \theta,\\
 F_\pm( {\bf w}_*) &=& \mp \left[ \ln \left( \gamma k_0 \sin \theta \right) + 1 \right], \\
 \mathcal M^{{\bf w},ab}_\pm &=& \pm ( 3 \gamma^2 \sin^2 \theta)^{-1} \delta^{ab}, \\
 |  \mathcal M^{{\bf w},ab}_\pm | &=& ( 3 \gamma^2 \sin^2 \theta)^{-3}, \\
 \mathcal M^{{\bf k},ab}_{\pm,ij} &=& \pm \gamma^2  \sin^2 \theta  \left[ 3  \delta_{ij} \delta^{ab} \right. \\
 && \left. \hspace{-0.9in} - 2 \left(  \cos \frac{2 \pi i}{3} {\hat {\bf x}} +  \sin \frac{2 \pi i}{3} {\hat {\bf y}} \right)^a \left( \cos \frac{2 \pi j}{3} {\hat {\bf x}} +  \sin \frac{2 \pi j}{3} {\hat {\bf y}}  \right)^b \right], \\
|  \mathcal M^{{\bf k},ab}_{\pm,ij} | &=& (3 \gamma^2 \sin^2 \theta)^9.
\end{eqnarray*}

The final answer is then very reminiscent of the power spectrum result (\ref{finalspectrum}) but with $N \rightarrow \sin \theta$,
\begin{eqnarray*}
\left( \frac{\Delta T}{T} \right)^3_{\rm osc} &=& B_0 \left( \frac{3H}{2 \pi M} \right)^3 e^{- 3(\gamma k_D \sin \theta)^{-2}} \\
&& \times \cos \left( \frac{M}{H} \left[ \ln (\gamma k_0 \sin \theta) + 1 \right] \right) . 
\end{eqnarray*}
So the bispectrum will also be suppressed at small $\theta$.

\section{Trispectrum}
\begin{figure}
\begin{center}
\hspace{-0.0in}
\includegraphics[scale=0.25]{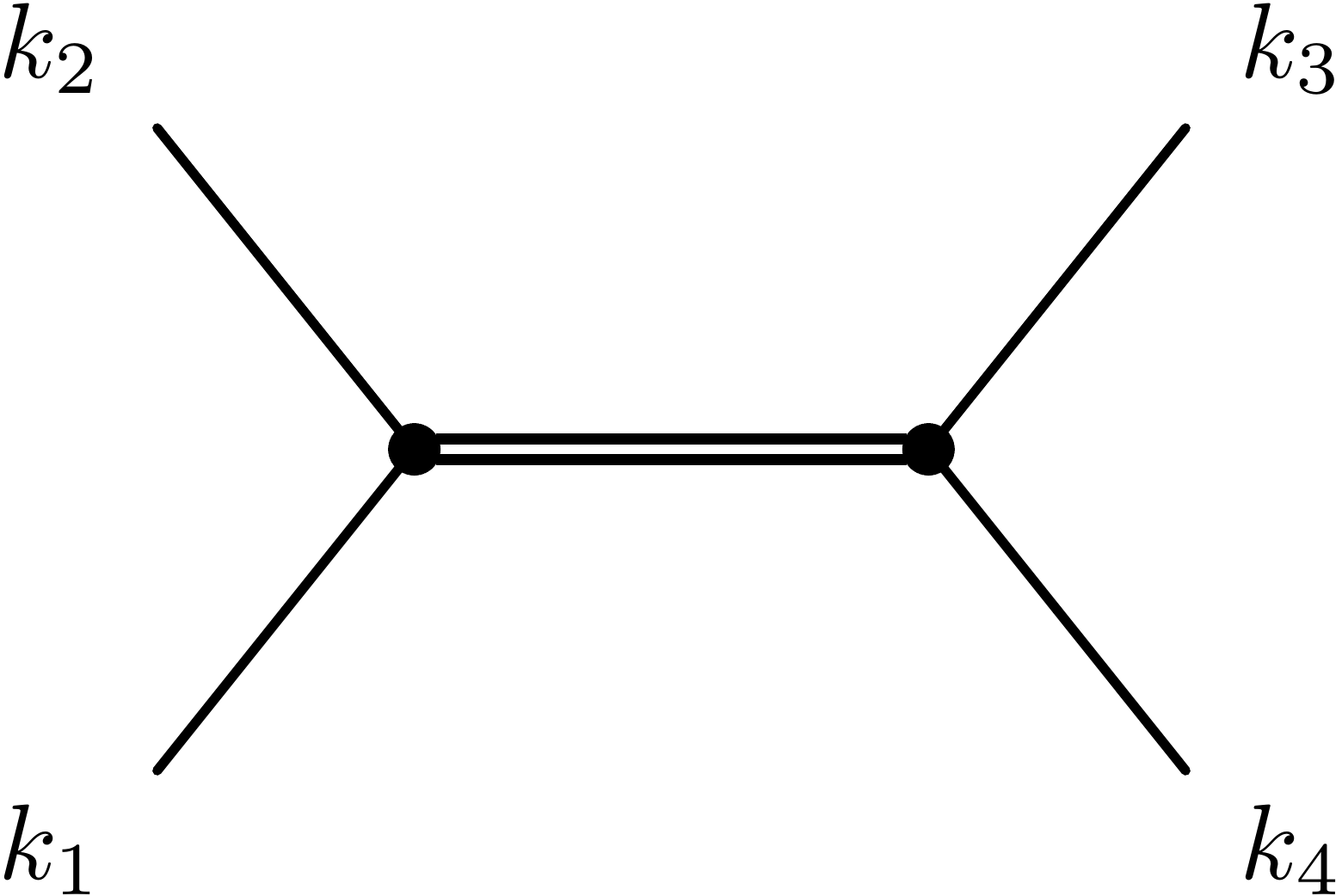} 
\caption{Tree-level high-energy contribution to the 4-point correlation function.  Single lines denote light fields, the double line denotes a heavy field.  The total correlation includes the two additional permutations of ${\bf k}_1,{\bf k}_2,{\bf k}_3,{\bf k}_4$.}
\label{tree_4pt_diags}
\end{center}
\end{figure}

The trispectrum is not as well-studied as the bispectrum but should soon also be an important tool in analyzing models of inflation \cite{Hu:2001fa, Kunz:2001ym, Fergusson:2010gn, Kehagias:2012td}.  The four-point correlation in position-space is
\begin{eqnarray}
\label{postri}
&& \hspace{-0.2in} \langle  \frac{\Delta T({\hat {\bf n}}_1)}{T}  \frac{\Delta T({\hat {\bf n}}_2)}{T}  \frac{\Delta T({\hat {\bf n}}_3)}{T}   \frac{\Delta T({\hat {\bf n}}_4)}{T}  \rangle \\
\nonumber
&=&  \int \prod_{i=1}^4 \frac{ d^3 {\bf k}_i}{(2 \pi)^3} T_{\rm rad}(k_i) \exp \left( - i \eta  {\bf k}_i  \cdot {\hat {\bf n}}_i \right)  \langle \varphi_{{\bf k}_1} \varphi_{{\bf k}_2} \varphi_{{\bf k}_3} \varphi_{{\bf k}_4} \rangle
\end{eqnarray}
where
\[  \langle \varphi_{{\bf k}_1} \varphi_{{\bf k}_2} \varphi_{{\bf k}_3} \varphi_{{\bf k}_4} \rangle = T_{\varphi} ({\bf k}_1,{\bf k}_2,{\bf k}_3,{\bf k}_4) (2 \pi)^3 \delta^3 \left( \sum_{i=1}^4 {\bf k}_i  \right). \]
The only theoretically motivated trispectrum oscillations were derived in \cite{Jackson:2012fu}, in which a light field in an excited state interacts with a heavy field, as shown in Figure~\ref{tree_4pt_diags} (here we are simplifying a bit to avoid subtleties of the ``in-in" formalism).  The correlation is given by
\begin{eqnarray}
\label{t}
T_ \varphi &=&  \frac{g^2 H^3}{2^{10} \pi^2(k_1 k_2 k_3 k_4)^{2} M} \times \\
\nonumber
&& \hspace{-0.5in} \left[ (k_1 + k_2)^2 - | {\bf k}_1 + {\bf k}_2|^2 \right]^{-1/4}  \left[ (k_3 + k_4)^2 - | {\bf k}_3 + {\bf k}_4|^2 \right]^{-1/4} \times \\
\nonumber
&& \hspace{-0.5in} \left( \frac{ k_1 + k_2 + \sqrt{ (k_1 + k_2)^2 - | {\bf k}_1 + {\bf k}_2|^2}}{k_3 + k_4 + \sqrt{ (k_3 + k_4)^2 - | {\bf k}_3 + {\bf k}_4|^2}} \right)^{-i M/H} + {\rm c.c.} \\
\nonumber
&& + {\rm permutations.}
\end{eqnarray}
This is scale invariant under $k_i \rightarrow \lambda k_i$.  Let us focus on the permutation written out, since the others are simple variations of these.  

In order to solve the stationary phase constraints, note that (aside from momentum-conservation) in the trispectrum correlation (\ref{t}) it is only ${\bf k}_1$ and ${\bf k}_2$ which are coupled only to each other, and likewise for ${\bf k}_3$ and ${\bf k}_4$:
\begin{eqnarray}
\label{tleft}
&& \hspace{-0.2in} T_\varphi (2 \pi)^3 \delta^3 \left( \sum_{i=1}^4 {\bf k}_i  \right) = \\
\nonumber
&&  \hspace{-0.3in}  \left( \frac{M}{H} \right)^3 \int d^3 {\bf w} \left[ T_\varphi^{{\bf k}_1,{\bf k}_2} e^{- \frac{M}{H}  i ({\bf k}_1+{\bf k}_2) \cdot {\bf w}} \right] \left[ T_\varphi^{{\bf k}_3,{\bf k}_4} e^{- \frac{M}{H} i ({\bf k}_3+{\bf k}_4) \cdot {\bf w}} \right]
\end{eqnarray}
where
\begin{eqnarray}
\nonumber
&& \hspace{-0.2in} T_\varphi^{{\bf k}_1,{\bf k}_2} =  \frac{g_1 H^{3/2} T_{\rm rad}(k_1) T_{\rm rad}(k_2) }{2^{5} \pi \sqrt{M} (k_1 k_2 )^{2} } \left[ (k_1 + k_2)^2 - | {\bf k}_1 + {\bf k}_2|^2 \right]^{-1/4} \\
\nonumber
&& \hspace{0.2in} \times  \left( k_1 + k_2 + \sqrt{2 k_1 k_2(1 - {\hat {\bf k}}_1 \cdot {\hat {\bf k}}_2)} \right)^{-i M/H}.
\end{eqnarray}
The $T_\varphi^{{\bf k}_3,{\bf k}_4}$ component is identical except for the opposite oscillation phase.  The factoring of (\ref{tleft}) allows tremendous simplification in solving the stationary phase constraints by expressing $F$ as the sum of a ``left" half containing only ${\bf k}_1$ and ${\bf k}_2$, and a ``right" half containing only ${\bf k}_3$ and ${\bf k}_4$, and finally also the prefactor $G$:
\begin{eqnarray*}
F( {\bf k}_i, {\bf w}) &=& F_L ({\bf k}_1,{\bf k}_2,{\bf w}) + F_R ({\bf k}_3,{\bf k}_4,{\bf w}), \\
G( {\bf k}_i) &=& G_L ( {\bf k}_1,{\bf k}_2) G_R ( {\bf k}_3,{\bf k}_4)
\end{eqnarray*}
where
\begin{eqnarray*}
G_L &\equiv& \frac{g H^{3/2} T_{\rm rad}(k_1) T_{\rm rad}(k_2) }{2^{5} \pi \sqrt{M} (k_1 k_2 )^{2}  \left[  2 k_1 k_2 \left( 1- {\hat {\bf k}}_1 \cdot {\hat {\bf k}}_2 \right)  \right]^{1/4}} , \\
F_L &\equiv& \ln \left( \frac{k_1 + k_2 + \sqrt{ 2 k_1 k_2 \left( 1- {\hat {\bf k}}_1 \cdot {\hat {\bf k}}_2 \right)} }{k_*} \right) \\
\nonumber
&+& \gamma \left( {\bf k}_1 \cdot {\hat {\bf n}_1} + {\bf k}_2 \cdot {\hat {\bf n}_2} \right) + \left( {\bf k}_1 + {\bf k}_2 \right) \cdot { {\bf w}} 
\end{eqnarray*}
and similarly for $G_R$ and $F_R$ with ${\bf k}_{1,2} \rightarrow {\bf k}_{3,4}$, and a sign difference in the logarithmic term of $F_R$.
\subsection{The First Half: Solving for ${\bf k}_1$, ${\bf k}_2$}
Taking the derivatives of $F_L$ with respect to ${\bf k}_1,{\bf k}_2$ results in the following extrema conditions:
\begin{eqnarray}
\label{ddk1}
- \frac{ {\hat {\bf k}}_1 +  \frac{  k_2 ( {\hat {\bf k}}_1 -  {\hat {\bf k}}_2) }{\sqrt{ 2 k_1 k_2 \left( 1- {\hat {\bf k}}_1 \cdot {\hat {\bf k}}_2 \right)}  }}{k_1 + k_2 + \sqrt{ 2 k_1 k_2 \left( 1- {\hat {\bf k}}_1 \cdot {\hat {\bf k}}_2 \right)}  } &=& \gamma {\hat {\bf n}_1} + {\bf w} , \ \ \\
\label{ddk2}
- \frac{ {\hat {\bf k}}_2 +  \frac{  k_1 ( {\hat {\bf k}}_2 -  {\hat {\bf k}}_1) }{\sqrt{ 2 k_1 k_2 \left( 1- {\hat {\bf k}}_1 \cdot {\hat {\bf k}}_2 \right)}  }}{k_1 + k_2 + \sqrt{ 2 k_1 k_2 \left( 1- {\hat {\bf k}}_1 \cdot {\hat {\bf k}}_2 \right)}  } &=& \gamma {\hat {\bf n}_2} + {\bf w} . \ \ 
\end{eqnarray}

In analyzing these equations, it will be helpful to make a change of variables:
\[ {\bf K} \equiv {\hat {\bf k}_1} - {\hat {\bf k}_2}, \hspace{0.5in} {\bf N} \equiv {\hat {\bf n}_1} - {\hat {\bf n}_2}, \]
\[ {\bf v} \equiv \frac{2}{\gamma} {\bf w} + {\hat {\bf n}_1} + {\hat {\bf n}_2}. \]
Rewriting (\ref{ddk1}) and (\ref{ddk2}) in these variables yields
\begin{eqnarray}
\label{ddk1new}
- \frac{ {\hat {\bf k}}_1 +  \frac{  k_2 {\bf K} }{K \sqrt{  k_1 k_2 }  }}{k_1 + k_2 + K \sqrt{  k_1 k_2}  } &=& \frac{\gamma}{2} \left( {\bf v}  + {\bf N} \right) , \ \ \\
\label{ddk2new}
- \frac{ {\hat {\bf k}}_2 -  \frac{  k_1 {\bf K} }{K \sqrt{  k_1 k_2 }  }}{k_1 + k_2 + K \sqrt{  k_1 k_2}  } &=& \frac{\gamma}{2} \left(  {\bf v} -{\bf N}\right)
\end{eqnarray}
where we have used $K = \sqrt{2(1 -  {\hat {\bf k}}_1 \cdot {\hat {\bf k}}_2)}$.  Dotting (\ref{ddk1new}) into (\ref{ddk2new}) gives
\begin{equation}
\label{dot1}
 \frac{ KN}{ k_1 + k_2 + K \sqrt{k_1 k_2}} = \frac{\gamma}{2} \left( N^2 - v^2 \right).
 \end{equation}
There are now two linear combinations of (\ref{ddk1new}), (\ref{ddk2new}) which will prove useful.  Subtracting,
\begin{equation}
- \frac{ {\bf K}}{K \sqrt{k_1 k_2}} =  \gamma {\bf N}. 
 \end{equation}
Squaring this gives
\begin{equation}
 \frac{1}{k_1 k_2} = \gamma^2 N^2 . 
 \end{equation}
Thus
\begin{equation}
\label{knunits}
{\hat {\bf K}} = - {\hat {\bf N}}. 
\end{equation} 
Now adding (\ref{ddk1new}) and (\ref{ddk2new}),
\begin{equation}
\label{kadd}
\frac{ - ( {\hat {\bf k}}_1 + {\hat {\bf k}}_2) + \frac{(k_1 - k_2) }{K \sqrt{k_1 k_2}} {\bf K} }{ k_1 + k_2 + K \sqrt{k_1 k_2}} = \gamma {\bf v}.
\end{equation} 

Since $( {\hat {\bf k}}_1 + {\hat {\bf k}}_2) \cdot {\bf K}$ = 0 it is clear that ${\bf v}$ should be decomposed according to
\[ {\bf v} = m {\hat {\bf M}} + n {\hat {\bf N}}  \]
where ${\hat {\bf M}}$ contains a $U(1)$ symmetry parameterized by the angle $\phi$,
\begin{eqnarray}
\nonumber
 {\hat {\bf M}}(\phi) &\equiv&  \cos \phi \frac{ {\hat {\bf n}}_1+ {\hat {\bf n}}_2 }{ \sqrt{4 - N^2}} + \sin \phi {\hat {\bf n}}_1 \times {\hat {\bf n}}_2  \\
 \label{kmunits}
 &=& - \frac{ {\hat {\bf k}}_1+ {\hat {\bf k}}_2 }{ \sqrt{4 - K^2}} .
\end{eqnarray}

Using (\ref{dot1}), the vector constraint (\ref{kadd}) can then be decomposed into the component along ${\hat {\bf M}}$,
\begin{equation}
\label{meq}
\frac{  \sqrt{4-K^2} }{2KN} \left( N^2 - m^2 - n^2  \right) = m
\end{equation}
and along ${\hat {\bf N}}$,
\begin{equation}
\label{neq}
- \frac{\gamma(k_1 - k_2)}{2K }  \left( N^2 - m^2 - n^2  \right) =  n. 
\end{equation}
Eq. (\ref{meq}) is then easily seen to be
\begin{eqnarray*}
m &=& \frac{2N}{\sqrt{4-K^2}} \sin \theta - \frac{NK}{\sqrt{4-K^2}}, \\
n &=&  \frac{2N}{\sqrt{4-K^2}}  \cos \theta.
\end{eqnarray*}
Note that because solutions to (\ref{dot1}) exist only for $v^2=m^2+n^2\leq N^2$, and (\ref{meq}) only for $m \geq 0$ we limit ourselves to a certain interval of $\varphi$,
\[ \varphi_{\rm max} \equiv \arcsin \frac{ K}{2} . \]
If we define
\[ y \equiv \frac{K}{\sqrt{4-K^2}}  \]
this can be inverted as
\[ K(r) = \frac{2r}{ \sqrt{1+r^2}} . \]
The solution for ${\bf v}$ can then be written as
\begin{eqnarray}
\label{vexpression}
{\bf v}(r,\varphi,\phi) &=& \left( N \sqrt{r^2+1} \sin \varphi - N r \right)  {\hat {\bf M}}(\phi) \\
\nonumber
&& \hspace{0.5in} +  N \sqrt{r^2+1} \cos \varphi {\hat {\bf N}},
\end{eqnarray}
where the range is given by (note that $r$ is not actually the radius, merely a convenient parameter)
\[ 0 \leq  \phi < 2 \pi, \hspace{0.2in} - \varphi_{\rm max} \leq \varphi \leq \varphi_{\rm max}, \hspace{0.2in} 0 \leq r < \infty  \]
where
\[ \varphi_{\rm max} = \arcsin \frac{ r}{\sqrt{r^2+1}} . \]
The solution to ${\bf v}$ thus fills a sphere of radius $N$.  Each $r$ represents an (American) football-shaped surface suspended between ${\bf v} = \pm {\bf N}$ corresponding to a contour of interaction time.  This can be seen by
\[ |\tau| = \frac{M}{H} \left( K \sqrt{k_1 k_2} \right)^{-1} = \frac{M \gamma N}{H K(r)} = \frac{\eta N}{K(r)} . \]
\begin{figure}
\begin{center}
\includegraphics[width=2.2in]{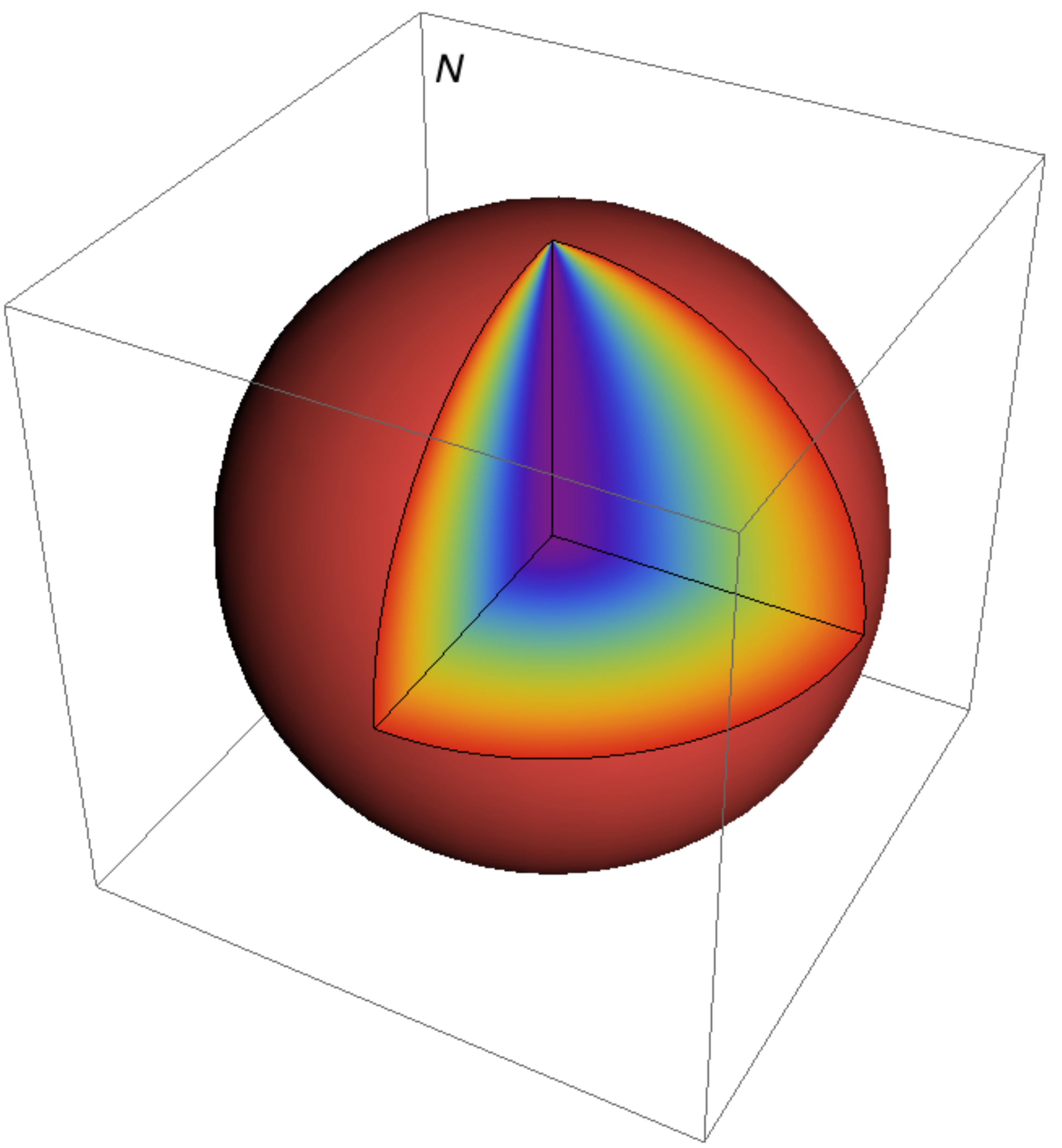}
\caption{Solutions to ${\bf v}$.  Each point in the sphere of radius $N$ corresponds to a unique solution of ${\bf k}_1,{\bf k}_2$.  The color indicates the time of interaction $\tau$.  The vertical axis is the direction ${\hat {\bf N}}$ and the transverse directions correspond to ${\hat {\bf M}}$.}
\label{vsolutions}
\end{center}
\end{figure}
Thus the position-space trispectrum at decoupling radius $\eta$ really does project onto only those interactions taking place at $|\tau| \sim \eta$.  Figure~\ref{vsolutions} shows the space of solutions to ${\bf v}$.

It is then easy to obtain
\begin{eqnarray*}
k_1 + k_2 &=& \frac{K}{\gamma} \left( \frac{2 N}{N^2-v^2} - \frac{1}{ N} \right), \\
k_1 - k_2 &=& - \frac{2 Kn}{\gamma( N^2-v^2)} .
\end{eqnarray*}
Then linear combinations produce
\begin{eqnarray}
\label{kmag}
k_1 &=& \frac{K}{2\gamma} \left[ \frac{2 N}{N^2-v^2}  \left( 1 - \frac{n}{N} \right) - \frac{1}{N} \right], \\
\nonumber
k_2 &=& \frac{K}{2\gamma} \left[ \frac{2 N}{N^2-v^2}  \left( 1 + \frac{n}{N} \right) - \frac{1}{N} \right].
\end{eqnarray}
Using (\ref{knunits}) and (\ref{kmunits}) we can also solve explicitly for the unit vectors,
\begin{eqnarray}
\label{kdir}
 {\hat {\bf k}}_1 &=& - \frac{1}{2} \left(  \sqrt{4-K(r)^2}  {\hat {\bf M}} + K(r){\hat {\bf N}} \right), \\
 \nonumber
 {\hat {\bf k}}_2 &=& - \frac{1}{2} \left(  \sqrt{4-K(r)^2}  {\hat {\bf M}} - K(r) {\hat {\bf N}} \right).
\end{eqnarray}
We can of course combine them to get ${\bf k}_1, {\bf k}_2$ but these are not so interesting by themselves.  More important is the combined momentum,
\begin{eqnarray}
\nonumber
  {\bf k}_1 +{\bf k}_2 &=& - \frac{\sqrt{4-K^2}}{2 \gamma} \left(\frac{2 KN}{N^2-v^2}  - \frac{K}{N} \right) {\hat {\bf M}} \\
  \nonumber
  && \hspace{0.3in} + \frac{K^2n}{\gamma(N^2-v^2)}  {\hat {\bf N}} \\
  \nonumber
&=& - \frac{ K^2 (N^2 + v^2)}{ \gamma (N^2-v^2)^2} {\bf v} + \frac{2(KN)^2 n}{\gamma (N^2-v^2)^2} {\hat {\bf N}} \\
\nonumber
&=&  \frac{4}{ \gamma \left[ (N^2+v^2)^2 - 4 n^2 N^2 \right] } \\
\label{k1k2}
&& \hspace{0.5in} \times \left[ - (N^2 + v^2) {\bf v} + 2 N n { {\bf N}} \right]
\end{eqnarray}
where we have used (\ref{meq}),
\[  K(r)^2 =  \frac{4 (N^2-v^2)^2}{(N^2+v^2)^2-4 n^2 N^2 } . \]

The fluctuations are 
\begin{equation}
\label{trifluct}
 \mathcal M^{L,ab}_{ij} \equiv \left. \frac{\partial^2 F_L}{\partial k^a_i \partial k^b_j} \right|_{k_*} 
\end{equation} 
which are given by
\begin{eqnarray*}
\mathcal M_{11}^{L,ab}  &=& \frac{ \left[  {\hat { k}}^a_1 +  \frac{k_2 ( {\hat { k}}^a_1 -  {\hat { k}}^a_2 ) }{\sqrt{ 2 k_1 k_2 (1 - \hat {\bf k}_1 \cdot \hat {\bf k}_2) }} \right]  \left[  {\hat { k}}^b_1 +  \frac{k_2 ( {\hat { k}}^b_1 -  {\hat { k}}^b_2 ) }{\sqrt{ 2 k_1 k_2 (1 - \hat {\bf k}_1 \cdot \hat {\bf k}_2) }} \right] }{ \left( k_1 + k_2 + \sqrt{ 2 k_1 k_2 (1 - \hat {\bf k}_1 \cdot \hat {\bf k}_2)} \right)^2} \\
&& \hspace{-0.4in} - \left. \frac{\frac{ \delta^{ab} -  {\hat { k}}^a_1 {\hat { k}}^b_1}{k_1} + \frac{\frac{k_2}{k_1} \left( \delta^{ab} - {\hat { k}}^a_1 {\hat { k}}^b_1 \right)}{\sqrt{ 2 k_1 k_2 (1 - \hat {\bf k}_1 \cdot \hat {\bf k}_2)} } - \frac{k_2^2 ( {\hat { k}}^a_1 -  {\hat { k}}^a_2 )( {\hat { k}}^b_1 -  {\hat { k}}^b_2 ) }{   \left(  2 k_1 k_2 (1 - \hat {\bf k}_1 \cdot \hat {\bf k}_2) \right)^{3/2} }}{ \left( k_1 + k_2 + \sqrt{ 2 k_1 k_2 (1 - \hat {\bf k}_1 \cdot \hat {\bf k}_2)} \right)} \right|_{k_*} , \\
\mathcal M_{12}^{L,ab} &=& \frac{ \left[  {\hat { k}}^a_1 +  \frac{k_2 ( {\hat { k}}^a_1 -  {\hat { k}}^a_2 ) }{\sqrt{ 2 k_1 k_2 (1 - \hat {\bf k}_1 \cdot \hat {\bf k}_2) }} \right]  \left[  {\hat { k}}^b_2 +  \frac{k_1 ( {\hat { k}}^b_2 -  {\hat { k}}^b_1 ) }{\sqrt{ 2 k_1 k_2 (1 - \hat {\bf k}_1 \cdot \hat {\bf k}_2) }} \right] }{ \left( k_1 + k_2 + \sqrt{ 2 k_1 k_2 (1 - \hat {\bf k}_1 \cdot \hat {\bf k}_2)} \right)^2} \\
&-& \left. \frac{ \frac{ {\hat { k}}^a_1 {\hat { k}}^b_2 - \delta^{ab} }{\sqrt{ 2 k_1 k_2 (1 - \hat {\bf k}_1 \cdot \hat {\bf k}_2)} } - \frac{k_1 k_2 ( {\hat { k}}^a_1 -  {\hat { k}}^a_2 )( {\hat { k}}^b_2 -  {\hat { k}}^b_1 ) }{   \left(  2 k_1 k_2 (1 - \hat {\bf k}_1 \cdot \hat {\bf k}_2) \right)^{3/2} }}{ \left( k_1 + k_2 + \sqrt{ 2 k_1 k_2 (1 - \hat {\bf k}_1 \cdot \hat {\bf k}_2)} \right)} \right|_{k_*}, \\
\end{eqnarray*}
where $|_{k_*}$ reminds us to evaluate the answer using the expressions (\ref{kmag}), (\ref{kdir}) in terms of ${\bf v}$ and ${\bf N}$.  Of course $\mathcal M_{22}^{L,ab}$ is analogous to $\mathcal M_{11}^{L,ab}$ with ${\bf k}_1 \leftrightarrow {\bf k}_2$.

\subsection{The Second Half: Solving for ${\bf k}_3$, ${\bf k}_4$}
The identical logic carries through for ${ {\bf k}}_{3,4}$, except there is an overall minus sign for the LHS of (\ref{ddk1}), (\ref{ddk2}) and so the solutions are simply the opposite of ${\hat {\bf k}}_{1,2}$:
\begin{eqnarray*}
 {\hat {\bf k}}_3 &=& \frac{1}{2} \left(  \sqrt{4-K(r_{34})^2}  {\hat {\bf M}}_{34} + K (r_{34}){\hat {\bf N}}_{34} \right), \\
 {\hat {\bf k}}_4 &=&  \frac{1}{2} \left(  \sqrt{4-K(r_{34})^2}  {\hat {\bf M}}_{34} - K(r_{34}) {\hat {\bf N}}_{34} \right).
\end{eqnarray*}
Here we have labeled $ {\bf N}_{34} \equiv {\hat {\bf n}}_3 - {\hat {\bf n}}_4$, as opposed to the original ${\bf N} \equiv {\bf N}_{12}$, and used the appropriate coordinate $r_{34}$.  The magnitudes $k_3,k_4$ are analogous to $k_1,k_2$, respectively.  The fluctuation matrix $\mathcal M^{R}$ is opposite to $\mathcal M^{L}$ given in (\ref{trifluct}), 
\[ \mathcal M^{R,ab}_{ij} = - \mathcal M^{L,ab}_{ij} \]
with ${\bf k}_{3,4}$ replacing ${\bf k}_{1,2}$.
\subsection{Solving the Complete System}
\begin{figure}
\begin{center}
\includegraphics[width=3in]{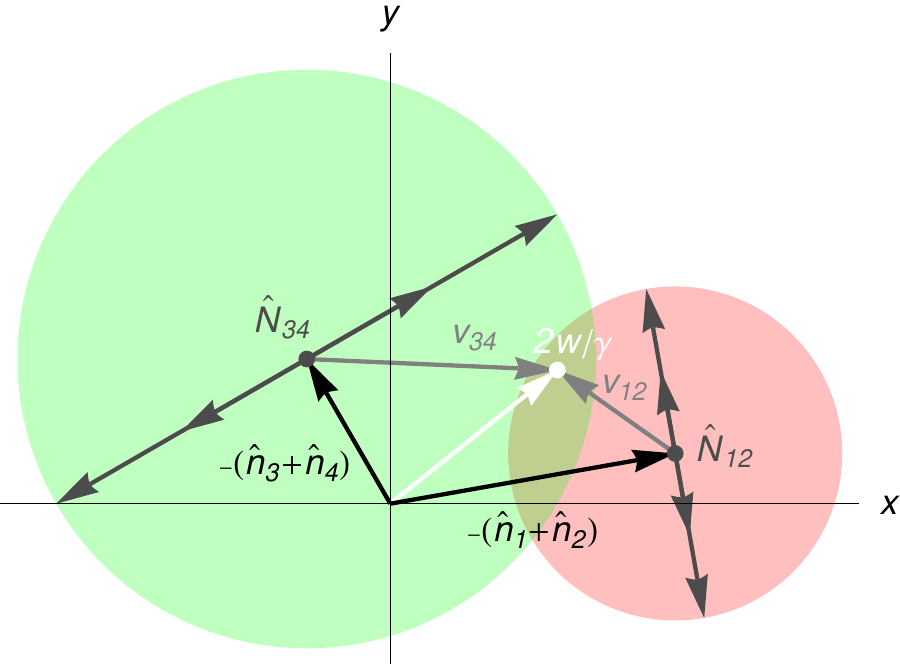}
\caption{Solution to the auxiliary variable equation of motion which must satisfy both halves of the system.}
\label{solution}
\end{center}
\end{figure}

At this point we have two separate sets of solutions: one for ${ {\bf k}}_{1,2}$ and one for ${ {\bf k}}_{3,4}$.  Substituting them back into the exponent function $F$ and returning ${\bf v}_{12},{\bf v}_{34}$ to the variable ${\bf w}$ gives
\begin{equation}
\label{fullaction}
 F( {\bf w}) = - \ln \frac{ \left| {\bf w} + \gamma {\hat {\bf n}}_1 \right| \left| {\bf w} + \gamma {\hat {\bf n}}_2 \right| }{ \left| {\bf w} + \gamma {\hat {\bf n}}_3 \right| \left| {\bf w} + \gamma {\hat {\bf n}}_4 \right| } . 
 \end{equation}
Taking $\partial F / \partial w^a$ yields the stationary phase constraint
\begin{equation}
\label{momcons}
0 = \frac{ w_*^a + \gamma {\hat { n}}^a_1}{\left| {\bf w}_* + \gamma {\hat {\bf n}}_1 \right|^2} + \frac{ {w}^a_* + \gamma {\hat { n}}^a_2}{\left| {\bf w}_* + \gamma {\hat {\bf n}}_2 \right|^2} -  \frac{ {w}^a_* + \gamma {\hat { n}}^a_3}{\left| {\bf w}_* + \gamma {\hat {\bf n}}_3 \right|^2} - \frac{ {w}^a_* + \gamma {\hat { n}}^a_4}{\left| {\bf w}_* + \gamma {\hat {\bf n}}_4 \right|^2} . 
\end{equation}
As in the bispectrum case, this constraint represents the conservation of momentum and could have been obtained by setting $\sum_i {\bf k}_i = 0$. 

We desire a solution to ${\bf w}$ which satisfies both the left and right halves, as shown in Figure~\ref{solution}.  Note that because ${\bf w}$ must be a solution to both halves, it must correspond to a ${\bf v}_{12}$ which must be both within the sphere of radius $N_{12}$ centered at $-( {\hat {\bf n}}_1 +  {\hat {\bf n}}_2)$ as well as a ${\bf v}_{34}$ within the sphere of radius $N_{34}$ centered at $-( {\hat {\bf n}}_3 +  {\hat {\bf n}}_4)$. 

To facilitate solution construction, a nice interpretation of the system can be made in terms of (logarithmic potential) electrostatics as follows.  Consider four charges, two positive and two negative, placed at locations ${\hat {\bf n}}_i$ on the unit sphere.  Then $F$ represents the potential of a test charge at location $- {\bf w}/\gamma$ in this background, and the condition (\ref{momcons}) is the condition for the test charge to be in equilibrium, stable or unstable.   

The fluctuation matrix for the auxiliary parameter is 
\begin{eqnarray*}
\mathcal M^{ {\bf w},ab} &\equiv& \left. \frac{\partial ^2 F}{\partial w^a \partial w^b} \right|_{ {\bf k}_*,{\bf w}_*} \\
&& \hspace{-0.8in} = \sum_{i=1}^4 \frac{q_i}{ | {\bf w}_* + \gamma {\hat {\bf n}}_i |^4} \left[ | {\bf w}_* + \gamma {\hat {\bf n}}_i |^2 \delta^{ab} -2 ( w^a_* + \gamma {\hat { n}}_i^a )( w^b_* + \gamma {\hat { n}}_i^b ) \right]
\end{eqnarray*}
where $q_{1,2} = +1$ and $q_{3,4}=-1$.  Performing the Gaussian integral, the final answer is
\begin{eqnarray*}
 \left( \frac{ \Delta T}{T} \right)^4_{\rm osc}  &=&  \left( \frac{H}{2 \pi M} \right)^{9/2}  G ( {\bf k}_*)e^{-i \frac{M}{H} F( {\bf k}_*,{\bf w}_*) } \\
 && \hspace{-0.2in} \times \frac{1}{ \sqrt{ \left| \mathcal M^L \right| \left| \mathcal M^R \right| \left| \mathcal M^{\bf w}  \right|}  } + {\rm c.c.} + {\rm permutations}.
 \end{eqnarray*}
 Finally we must include the contribution of the flat power spectrum.  This amounts to simply summing the permutations of the power spectrum-squared,
\begin{eqnarray*}
\left( \frac{ \Delta T}{T} \right)^4_{\rm gaussian} &=& \left( \frac{ \Delta T}{T} \right)^2_{N_{12}} \left( \frac{ \Delta T}{T} \right)^2_{N_{34}} \\
&& \hspace{-0.5in} + \left( \frac{ \Delta T}{T} \right)^2_{N_{13}} \left( \frac{ \Delta T}{T} \right)^2_{N_{24}} + \left( \frac{ \Delta T}{T} \right)^2_{N_{14}}\left( \frac{ \Delta T}{T} \right)^2_{N_{23}}. 
\end{eqnarray*}

\subsection{A Family of Solutions}
While a general solution to (\ref{momcons}) appears to be difficult, there is one family of solutions which is obvious.   In the language of the electrostatics analogy, place the two positive charges on the unit sphere at ${\hat {\bf n}}_{1,2}$ around the ${\hat {\bf z}}$-axis symmetrically at angle $\theta$ so that they are separated by length $N$, and use their difference vector ${\bf N}_{12} \equiv {\hat {\bf n}}_{1} - {\hat {\bf n}}_{2}$ to define ${\hat {\bf x}}$.  Now place the negative charges at ${\hat {\bf n}}_{3,4}$ at the same height, but rotated by some angle $\phi$ with respect to ${\hat {\bf x}}$.  This makes the directional unit vectors
\begin{eqnarray}
\label{trifamily}
{\hat {\bf n}}_1 &=& \sin \theta {\hat {\bf x}} + \cos \theta {\hat {\bf z}}, \\
\nonumber
{\hat {\bf n}}_2 &=& - \sin \theta {\hat {\bf x}} + \cos \theta {\hat {\bf z}}, \\
\nonumber
{\hat {\bf n}}_3 &=& \sin \theta ( \cos \phi {\hat {\bf x}} + \sin \phi {\hat {\bf y}}) + \cos \theta {\hat {\bf z}}, \\
\nonumber
{\hat {\bf n}}_4 &=&  - \sin \theta ( \cos \phi {\hat {\bf x}} + \sin \phi {\hat {\bf y}}) + \cos \theta {\hat {\bf z}}, \\
\nonumber
{\bf N}_{12} &=& 2 \sin \theta  {\hat {\bf x}} , \\
\nonumber
{\bf N}_{34} &=& 2 \sin \theta ( \cos \phi {\hat {\bf x}} +  \sin \phi {\hat {\bf y}} ), \\
\nonumber
{\hat {\bf M}}_{12} &=& - {\hat {\bf M}}_{34} = {\hat {\bf z}} .
\end{eqnarray}
\begin{figure}
\begin{center}
\includegraphics[width=3in]{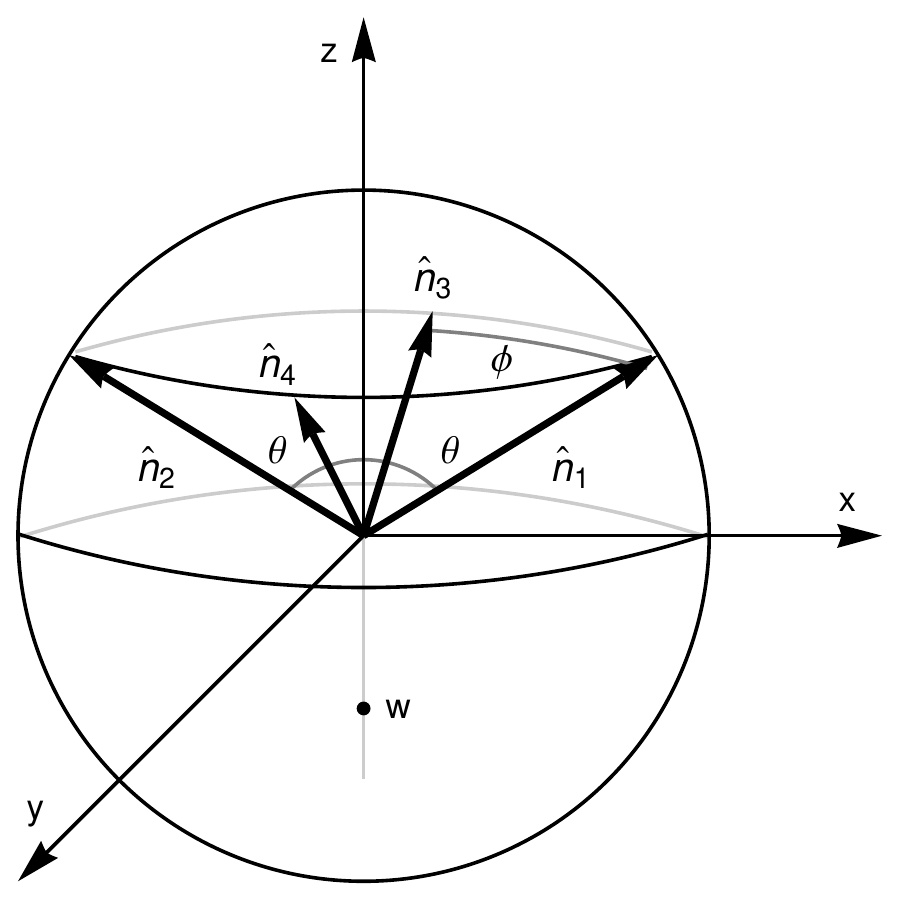}
\caption{A family of solutions to the trispectrum stationary phase constraints given by eq. (\ref{trifamily}).}
\label{ntri}
\end{center}
\end{figure}

\begin{figure}
\begin{center}
\includegraphics[width=2in]{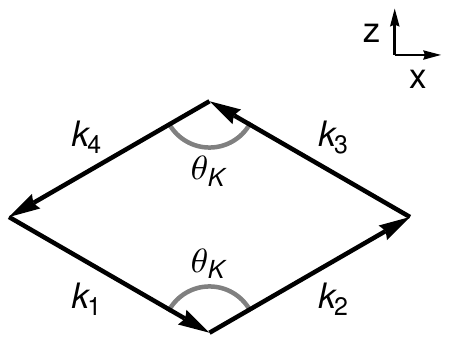} \\
\includegraphics[width=2in]{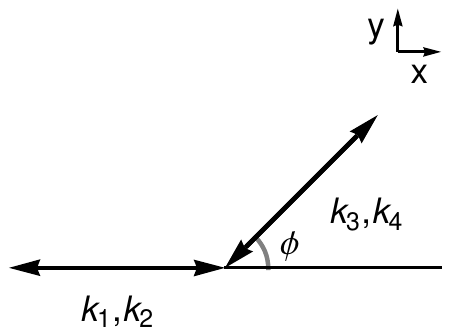}
\caption{The $k$-space configuration corresponding to the same family of solutions.}
\label{ktri}
\end{center}
\end{figure}
This is shown in Figure~\ref{ntri}.  Clearly a test charge located at any point on the ${\hat {\bf z}}$-axis will be in (unstable) equilibrium, the force from the positive charges canceling that from the negative ones.  Thus we consider solutions of the form
\[ {\bf w}_* = w^z {\hat {\bf z}}. \]
The fact that $v^2 \leq N^2$ means the solution must be within the sphere center located at
\[-({\hat {\bf n}}_1+{\hat {\bf n}}_2)=-({\hat {\bf n}}_3+{\hat {\bf n}}_4) = - 2 \cos \theta {\hat {\bf z}} \]
and have the bounds
\[  - \sin \theta \leq  \frac{w^z}{\gamma} + \cos \theta \leq \sin \theta . \]
Denoting
\[ \sigma \equiv \sin^2 \theta + \left( w^z/\gamma + \cos \theta \right)^2 \]
and $\mathcal M ^{' {\bf w}}$ as the fluctuation matrix without the ${\bf z}$-component,
we have the solution shown in Figure~\ref{ktri},

\begin{eqnarray*}
| {\bf w} + \gamma {\hat {\bf n}}_i |^2 &=& \gamma^2 \sigma, \\
 n_{12} &=& n_{34} = 0, \\
 m_{12} &=& - m_{34} = 2 \left( w^z / \gamma + \cos \theta \right), \\
r &=& \frac{ \sin \theta \left[ 1 - \left( \frac{w^z/\gamma + \cos \theta}{\sin \theta} \right)^2 \right]}{2 ( w^z / \gamma + \cos \theta)}, \\
K &=& \frac{2}{\sigma} \left[ \sin^2 \theta - \left( w^z/\gamma + \cos \theta \right)^2 \right], \\
\theta_{K} &=& \arccos \left( 1- \frac{K^2}{2} \right), \\
\left| \mathcal M^{\bf ' w} \right| &=& \frac{16 \sin^4 \theta \sin ^4 \phi}{ \gamma^4 \sigma^4 }, \\
F( {\bf w}_*) &=& 0, \\
G( {\bf w}_*) &=& \frac{2g^2 H^3 \left( \gamma \sin \theta \right)^{9} e^{- (\gamma k_D \sin \theta)^2} }{ (2 \pi)^2 M K}, \\
k_i &=& \frac{1}{2 \gamma \sin \theta}, \\
{\hat {\bf k}}_1 &=& - \frac{K}{2} {\hat {\bf x}} - \sqrt{1 - \left( \frac{K}{2} \right)^2}  {\hat {\bf z}} , \\
{\hat {\bf k}}_2 &=&  \frac{K}{2} {\hat {\bf x}} - \sqrt{1 - \left( \frac{K}{2} \right)^2}  {\hat {\bf z}} , \\
{\hat {\bf k}}_3 &=&  \frac{K}{2} \cos \phi {\hat {\bf x}} + \frac{K}{2} \sin \phi {\hat {\bf y}} + \sqrt{1 - \left( \frac{K}{2} \right)^2}  {\hat {\bf z}} , \\
{\hat {\bf k}}_3 &=& - \frac{K}{2} \cos \phi {\hat {\bf x}} - \frac{K}{2} \sin \phi {\hat {\bf y}} + \sqrt{1 - \left( \frac{K}{2} \right)^2}  {\hat {\bf z}} .
\end{eqnarray*}

The fluctuation matrix components are
\begin{eqnarray*}
\mathcal M^{L,ab}_{11} &=& \left( \frac{ \gamma \sigma}{2 \sin \theta } \right)^2 \left( {\hat k}_1^a - {\hat x}^a \right)  \left( {\hat k}_1^b - {\hat x}^b \right) \\
&-& \gamma^2 \sigma \left[ \left( \delta^{ab} - {\hat k}_1^a {\hat k}_1^b \right) \left( 1 + \frac{1}{K} \right) - \frac{1}{K} {\hat x}^a {\hat x}^b\right] , \\
\mathcal M^{L,ab}_{12} &=&  \mathcal M^{L,ba}_{21} = \left( \frac{ \gamma \sigma}{2 \sin \theta } \right)^2 \left( {\hat k}_1^a - {\hat x}^a \right)  \left( {\hat k}_2^b + {\hat x}^b \right) \\
&+& \gamma^2 \sigma \left[ \left( \delta^{ab} - {\hat k}_1^a {\hat k}_2^b \right)  - \frac{1}{K} {\hat x}^a {\hat x}^b\right], \\
\mathcal M^{L,ab}_{22} &=& \left( \frac{ \gamma \sigma}{2 \sin \theta } \right)^2 \left( {\hat k}_2^a + {\hat x}^a \right)  \left( {\hat k}_2^b + {\hat x}^b \right) \\
&-& \gamma^2 \sigma \left[ \left( \delta^{ab} - {\hat k}_2^a {\hat k}_2^b \right) \left( 1 + \frac{1}{K} \right) - \frac{1}{K} {\hat x}^a {\hat x}^b\right] . \\
\end{eqnarray*}
Since $\mathcal M_{ij}^{R,ab} = - \mathcal M_{ij}^{L,ab}$, the magnitudes of their determinants must be equal, $| \mathcal M^R | = | \mathcal M^L |$.

Evaluation of the fluctuation matrix determinants is too involved analytically, and so we resort to numerical methods.
Assembling everything together, and remembering that $w^z$ is unconstrained, the correlation will be
\begin{eqnarray}
\label{trianswer}
\left( \frac{ \Delta T}{T} \right)^4_{\rm osc} &=& \left( \frac{H}{2 \pi M} \right)^4 \\
\nonumber
&& \hspace{-0.4in} \times  \int_{- \gamma( \sin \theta + \cos \theta)}^{ \gamma( \sin \theta - \cos \theta)} \frac{dw^z}{2 \pi} \frac{G}{ \sqrt{ \left| \mathcal M^{L} \right|  \left| \mathcal M^{R} \right|  \left| \mathcal M^{\prime {\bf w}} \right|}} . 
\end{eqnarray}
We should also include the contributions from permuting the ${\bf k}_i$ in the interaction diagram shown in Figure~\ref{tree_4pt_diags}.  This is accomplished merely by permuting the directional vectors ${\hat {\bf n}}_i$.  Consider again the electrostatics analogy presented earlier in which ${\hat {\bf n}}_1$, ${\hat {\bf n}}_2$ denote the location of positive charges on the unit sphere whereas ${\hat {\bf n}}_3$, ${\hat {\bf n}}_4$ are the location of negative charges.  For the family of solutions under consideration this yielded equilibrium points for a test charge located along the $z$-axis.  If we now permute the charges so that ${\hat {\bf n}}_1$, ${\hat {\bf n}}_3$ are positive charges and ${\hat {\bf n}}_2$, ${\hat {\bf n}}_4$ are negative, simple electrostatic intuition shows that there will cease to be any equilibrium points (except in the special cases in which the ${\hat {\bf n}}_i$ overlap, which are singular anyway).  Thus there is no soution to the stationary phase equation, and the contribution vanishes to leading order.  The same holds true for the final permutation of ${\hat {\bf n}}_1$, ${\hat {\bf n}}_4$ being positive charges and ${\hat {\bf n}}_2$, ${\hat {\bf n}}_3$ being negative.  Thus we do not need to modify the answer (\ref{trianswer}).

The gaussian trispectrum result is given by
\begin{eqnarray*}
\left( \frac{ \Delta T}{T} \right)^4_{\rm gaussian} &=& \left[ \left( \frac{ \Delta T}{T} \right)^2_{N=2 \sin \theta} \right]^2 \\
&& \hspace{-1.1in} + \left[ \left( \frac{ \Delta T}{T} \right)^2_{N=  \sqrt{2(1- \cos \phi)}\sin \theta } \right]^2 +  \left[ \left( \frac{ \Delta T}{T} \right)^2_{N= \sqrt{2(1+ \cos \phi)} \sin \theta}  \right]^2 . 
\end{eqnarray*}

\begin{figure}
\begin{center}
\includegraphics[width=3in]{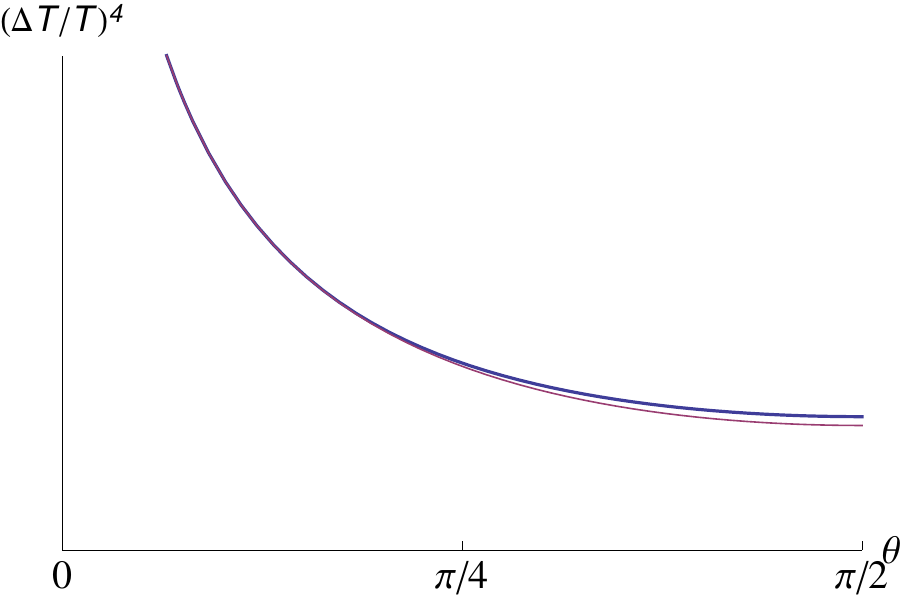}
\caption{Simplified CMB trispectrum in direction-space, varying $\theta$ while keeping fixed $\phi=\pi/2$.  The flat power spectrum is given by the thin red line.}
\label{tri_theta}
\end{center}
\end{figure}

\begin{figure}
\begin{center}
\includegraphics[width=3in]{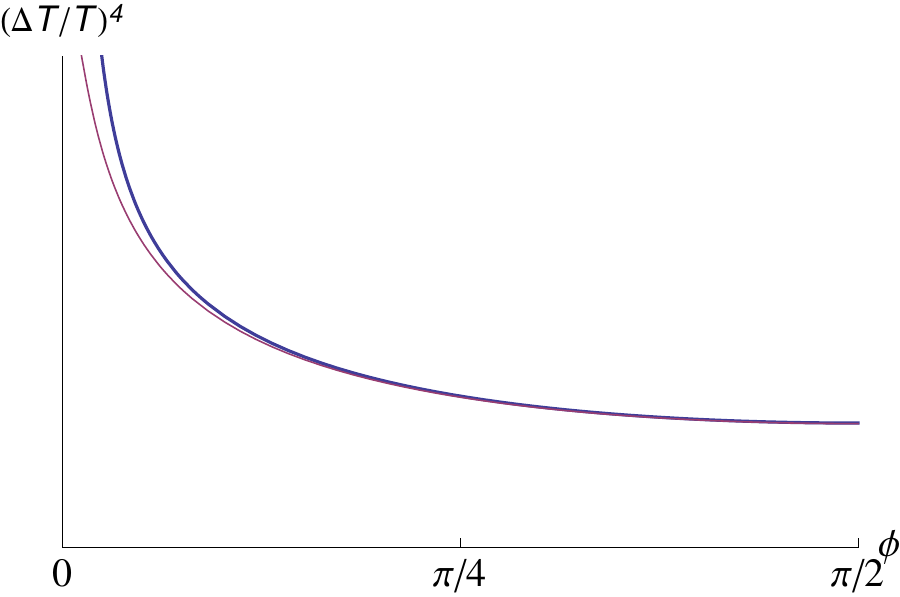}
\caption{Simplified CMB trispectrum in direction-space, varying $\phi$ while keeping fixed $\theta=\pi/2$.  The flat power spectrum is given by the thin red line.}
\label{tri_phi}
\end{center}
\end{figure}

In Figure~\ref{tri_theta} the trispectrum is shown as a function of $\theta$, keeping fixed $\phi=\pi/2$.  In Figure~\ref{tri_phi} this is evaluated as a function of $\phi$, keeping fixed $\theta=\pi/2$.  


\section{Discussion}
We calculated the 2-, 3-, and 4-point angular correlation functions for models with logarithmic oscillations, and found that they are essentially featureless.  Thus the direction-space representation does not appear to offer any advantages in detecting their presence compared to the conventional $k$-space analysis.  

Consider instead the power spectrum arising from a boundary effective field theory \cite{Greene:2005aj}, meaning the oscillations are periodic as a linear function of $k$,
\[ \Delta P_\varphi = \mathcal A k \cos \mathcal C k. \]
Fourier transforming this yields peaked correlations localized to definite angular separations,
\[ \left( \frac{ \Delta T}{T} \right)^2_{\rm osc} \sim \mathcal A \delta \left(  \mathcal C - N \eta \right) . \]
These have the advantage of only requiring certain configurations of angular separation, and thus are vastly more computationally efficient.  This would also work well for the bouncing cosmology model studied in \cite{Falciano:2008gt}, which obtains a complete power spectrum (not just a correction) of the form
\[ P_\varphi = \mathcal A k^{n_s-1} \cos^2 \left( \omega \frac{k}{k_0} + \varphi_0 \right). \]
For higher-point correlations this is even more efficient.  Suppose the bispectrum were of the form
\[ B_\varphi \sim \frac{B_0}{k_1^2 k_2^2 k_2^3} \cos \left[ \mathcal C (k_1 + k_2 + k_3) \right] . \]
The position-space correlation would then peak when 
\[ \left( \frac{\Delta T}{T} \right)^3 \sim B_0 \int \prod_i d^2 {\hat {\bf k}}_i  \ \delta \left( \mathcal C - \eta {\hat {\bf k}}_i \cdot  {\hat {\bf n}}_i \right) . \]
The solution to this forces the ${\hat {\bf k}}_i$ to take the same relative configuration as the ${\hat {\bf n}}_i$ but rotated an overall angle away, reducing this to a one-parameter family of solutions to examine.

Similar simplifications for the linear oscillations may be possible if modified transforms are used, such as a transformation from $\ell$-space back to angular scales with a logarithmic scale:
\[ P( \theta) \equiv \sum_\ell \frac{2\ell+1}{4 \pi} P_{\ln \ell} (\cos \theta) \mathcal C_\ell \]
with $P_{\ln \ell} (\cos \theta)$ the Legendre polynomial.  Oscillations which are periodic in the logarithm of $k$ such as (\ref{pphi}) will then appear as peaks at $\theta \approx N$,
\[ P_{\rm osc}( \theta)  \sim \beta \delta \left( \theta \eta - \frac{M}{H} \right) . \]
Such alternative search strategies merit further investigation.

\section{Acknowledgements}
We would like to thank R.~Easther, P.~D.~Meerburg, P.~Peter, C.~Ringeval, and K.~Schalm for helpful discussions.  This work benefitted from discussions at the Leiden Center Workshop ``Effective Field Theory in Inflation" in July 2012.

\end{document}